\documentclass{aa}  
\usepackage{graphicx}
\usepackage{txfonts}
\begin{document}
   \title{A new modified-rate approach for gas-grain chemical simulations}

   \author{R. T. Garrod\inst{1}
          }

   \offprints{R. T. Garrod}

   \institute{Max-Planck-Institut f{\"u}r Radioastronomie, 
   Auf dem H{\"u}gel 69, Bonn, 53121, Germany\\
              \email{rgarrod@mpifr-bonn.mpg.de}
             }

   \date{}

 
  \abstract
   {Understanding grain-surface processes is crucial to interpreting the chemistry in many regions of the interstellar medium. However, accurate surface chemistry models are computationally expensive and are difficult to integrate with gas-phase simulations.}
   {A new modified-rate method for solving grain-surface chemical systems is presented. The purpose of the method is to trade a small amount of accuracy, and certain excessive detail, for the ability to accurately model highly complex systems that can otherwise only be treated using the sometimes inadequate rate-equation approach.}
   {In contrast to previous rate-modification techniques, the functional form of the surface production rates was modified, and not simply the rate coefficient. This form is appropriate to the extreme ``small-grain'' limit, and can be verified using an analytical master-equation approach. Various further modifications were made to this basic form, to account for competition between processes, to improve estimates of surface occupation probabilities, and to allow a switch-over to the normal rate equations where these are applicable.}
   {The new method was tested against a number of systems solved previously using master-equation and Monte Carlo techniques. It is found that even the simplest method is quite accurate, and a great improvement over rate equations. Further modifications allow the master-equation results to be reproduced exactly for the methanol-producing system, within computational accuracy. Small discrepancies arise when non-zero activation energies are assumed for the methanol system, which result from complex reaction-competition processes that cannot be resolved easily without using exact methods. Inaccuracies in computed abundances are never greater than a few tens of percent, and typically of the order of one percent, in the most complex systems tested.}
   {The new modified-rate approach presented here is robust to a range of grain-surface parameters, and accurately reproduces the results of exact methods. Furthermore, it may be derived from basic approximations, making the behaviour of the system understandable in terms of physical processes rather than time-dependent probabilities or other more abstract quantities. The method is simple enough to be easily incorporated into a full gas-grain chemical code. Implementation of the method in simple networks, including hydrogen-only systems, is trivial, whilst the results are highly accurate.}


   \maketitle

\section{Introduction}

Chemical models consisting of hundreds of species and thousands of reactions are frequently used to interpret the morphologies and evolutionary histories of interstellar clouds and star-formation regions. Towards such ends, simulations dealing solely with gas-phase chemistry have achieved much success; however, the explicit consideration of grain-surface processes in chemical models is becoming increasingly important to our understanding of interstellar chemistry. From the formation of the most basic and abundant interstellar molecule, H$_2$ (e.g. Gould \& Salpeter 1963; Hollenbach \& Salpeter 1971; Duley \& Williams 1984), to some of the most chemically complex organic species observed in star-forming regions (e.g. Charnley et al. 1992, 1995; Cazaux et al. 2003; Horn et al. 2004; Garrod \& Herbst 2006; Garrod et al. 2008), the action of grain-surface chemical processes appears to be crucial. 

Unfortunately, the accurate coupling of gas-phase and grain-surface processes in chemical models is difficult, due to the different nature of the chemistry in each phase. Gas-phase chemistry may be accurately modelled by describing chemical abundances as averaged concentrations, or number densities. By assuming an arbitrarily large ``cell'' of gas, the absolute number of particles of any species present in the system can be assumed to be statistically large, making stochastic fluctuations unimportant. Such a ``deterministic'' treatment allows the chemistry to be described by a set of ordinary, first-order differential equations (rate equations), which are solved numerically, allowing the time-dependent behaviour to be traced.
 
The simplicity of this rate-equation approach naturally encourages its use in solving grain-surface problems, especially in coupled gas-phase/grain-surface (i.e. gas-grain) models; however, its accuracy in these applications is in some cases questionable. Grain-surface chemistry takes place on finite surfaces, where the populations of certain chemical species can become very small, of the order of 1. If surface reactions occur quickly in such a regime, then stochastic effects may come into play; this renders the rate-equation treatment inaccurate, yielding production rates that are faster than physically possible. 

These failures do not arise in all regimes; \cite{katz} showed that hydrogen diffusion at low temperature takes place via thermal hopping, rather than fast quantum tunnelling. For grains of canonical size ($0.1\mu$m), at temperatures of $\sim$10 K, the reaction rates are typically no faster than the accretion or evaporation rates of the reactants, putting the system into the deterministic limit. It is in the consideration of grains that are smaller or hotter than this that stochastic effects must be considered.

Monte Carlo methods have been developed (e.g. Tielens \& Hagen 1982; Charnley et al. 1997; Charnley 1998, 2001) to take account of stochastic behaviour. The most recent techniques (Cuppen \& Herbst 2005) trace the behaviour of individual atoms and molecules on a grain surface; however, even the simplest of such schemes is difficult to integrate with a rate equation-based gas-phase code (Chang et al. 2007). Alternatively, the master-equation method describes the system using the time derivatives of the probabilities of specific population states, which are easily solved in tandem with gas-phase rate equations (Biham et al. 2001). Cut-offs are imposed, representing the maximum populations of surface species. Unfortunately, with large networks the number of possible population-state combinations becomes unmanageable, even with low cut-offs. 

Stantcheva et al. (2002) developed hybrid schemes that mix deterministic and stochastic methods, treating abundant, ``deterministic'' species such as CO using rate-equations, but treating ``stochastic'' species using a master-equation method that employs low cut-offs. These schemes are successful for the small networks upon which they have been tested, but require prior knowledge from the exact methods in order to distinguish ``deterministic'' from ``stochastic'' species. 

Green et al. (2001) made approximations within the master-equation method to give analytic expressions for production rates in some simple chemical systems; but the approximations are valid only in the stochastic regime, ruling out an extension to complex networks involving deterministic species such as CO.

The ``method of moments'' recently developed by Barzel \& Biham (2007a,b) involves the calculation of second moments of population, beginning from a master-equation standpoint. This allows the calculation of accurate production rates. However, in the deterministic regime, the method gives accurate production rates, but can produce inaccurate populations (Barzel \& Biham, 2007b; B. Barzel, private comm.). It is currently unclear how this would affect the behaviour of large chemical networks. Also, the scheme is untested against networks that include activation energies, such as are necessary in the methanol-producing system.

Attempts have been made previously (Caselli et al. 1998; Shalabiea et al. 1998; Stantcheva et al. 2001) to rectify the inaccuracies of the rate-equation method by modifying the rate coefficients of grain-surface production rates according to a set of simple rules; \cite{caselli02} also devised a more complex modification scheme for reactions involving activation energy barriers. These modifications were largely empirical, and were not universally successful; convergence with exact techniques was achieved only for low temperatures, where atomic hydrogen is the only mobile reactant. The comparison of techniques conducted by Rae et al. (2003), using a standardised two-reactant system, found that the method was in general no more accurate than the standard rate-equation approach.

Here is presented a new method of modification of the production rates, different from previous techniques. This method modifies not the rate {\em coefficient}, but the functional form of the production rate itself, adopting production rates derived for the extreme ``small-grain'' regime. The method switches smoothly to a rate-equation treatment where appropriate. The scheme retains its functional dependence on averaged population values, $\langle N \rangle$, rather than analysing individual states, $N$, allowing it to be easily implemented in standard gas-grain codes. The purpose of the new method is not primarily to produce a {\em perfect} match to the results of exact methods, but to improve accuracy in large chemical networks that cannot be modelled with exact techniques.

Section 2 outlines the rates for surface processes. Basic modified rates are formulated in Section 3. Sections 4 and 5 detail further modifications to the basic rates. Section 6 applies the method to reaction systems with activation energies. A discussion follows in Section 7, and conclusions in Section 8.

\section{Accretion, Evaporation and Reaction Rates}

In the systems to be analysed in this paper, three basic grain-surface processes are considered: accretion of gas-phase atoms onto the grains; evaporation of grain-surface species into the gas phase; and reaction between species on the grain surfaces.

Here, the accretion rate is defined simply as an average flux of atoms/molecules per second. This obviates the need to trace any kind of gas-phase chemistry in the following tests. 

The evaporation rate of species $A$ is governed by the lifetime against evaporation of one particle, $t_{evap}(A)$. Assuming a binding energy $E_{D}(A)$ (in Kelvin), and a grain temperature $T_d$, the rate coefficient for evaporation is:
\begin{equation}
k_{evap}(A) = \nu \cdot \exp{ \left[ - E_{D}(A)/T_{d} \right]} = t_{evap}^{-1}(A)
\end{equation}
where $\nu$ is the vibrational frequency of the atom/molecule in its binding site, typically of the order of $10^{12}$ s$^{-1}$ (e.g., Hasegawa et al. 1992). Hence the total evaporation rate (in s$^{-1}$) is:
\begin{equation}
R_{evap}(A) = k_{evap}(A) \cdot \langle N(A) \rangle
\end{equation}
where $\langle N(A) \rangle$ is the average surface population of species $A$.

It is assumed that surface reactions take place via the Langmuir-Hinshelwood mechanism: reaction occurs when two reactants diffusing over the grain surface meet in the same binding site. The rate at which reaction occurs is:
 \begin{equation}
k_{AB} = \frac{ R_{hop}(A) + R_{hop}(B) }{S} \kappa _{AB}
\end{equation}
where $R_{hop}(A)$ and $R_{hop}(B)$ are the average rates of thermal hopping between binding sites for each reactant, $S$ is the number of binding sites on the grain, and $\kappa _{AB}$ is an efficiency factor that accounts for any activation energy required to react. If diffusion is assumed to occur via thermal hopping, the rate is simply
\begin{equation}
R_{hop}(A) = \nu \cdot \exp{ \left[ - E_{b}(A)/T_{d} \right]}
\end{equation}
where $E_{b}$ is the strength of the barrier (in Kelvin) between binding sites. If, in the case of atomic hydrogen, movement between sites is assumed to occur through quantum tunnelling, different expressions pertain; see \cite{hasegawa}.

Where an activation energy is required for reaction to occur, the efficiency, $\kappa _{AB}$, is typically described by a Boltzmann factor, or an expression for quantum-tunnelling efficiency, whichever is the more efficient. Alternatives to these simple expressions have been put forward that take account of competition between reaction and diffusion out of the binding site (Awad et al. 2005, Chang et al. 2007); however, for the purposes of comparison to older results, these more recent developments are ignored.

Using rate equations to solve a grain-surface chemical network, the production rate (in s$^{-1}$) of the products of the reaction between species $A$ and $B$ is defined as:
\begin{equation}
R_{prod}(AB) = k_{AB} \cdot \langle N(A) \rangle \cdot \langle N(B) \rangle
\end{equation}
or, in the case of homogeneous reactants,
\begin{equation}
R_{prod}(AA) = \frac{k_{AA}}{2} \cdot \langle N(A) \rangle ^{2} .
\end{equation}
As stated above, the accuracy of this definition breaks down when stochastic effects become important.

The difference between the production rate, $R_{prod}(AB)$, and the reaction rate, $k_{AB}$, should be noted. The former is the subject of the modifications proposed in this paper; the latter is strictly defined above in equation (3), and is not adjusted in any way.

\begin{figure}
\centering
\includegraphics[width=8cm]{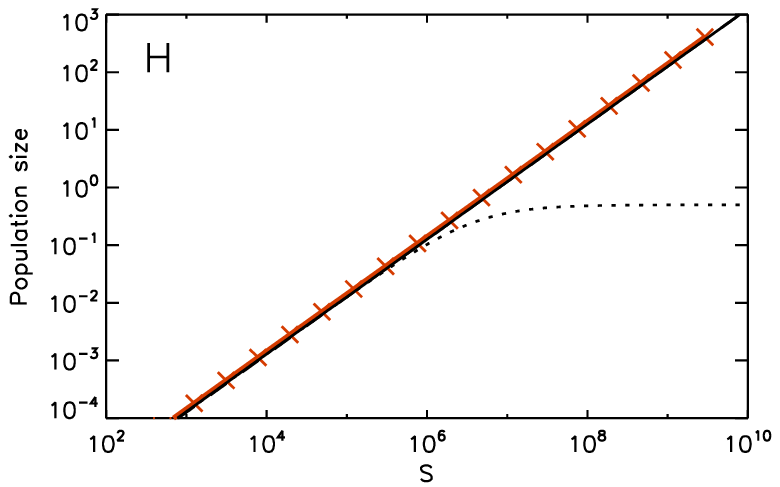}
\includegraphics[width=8cm]{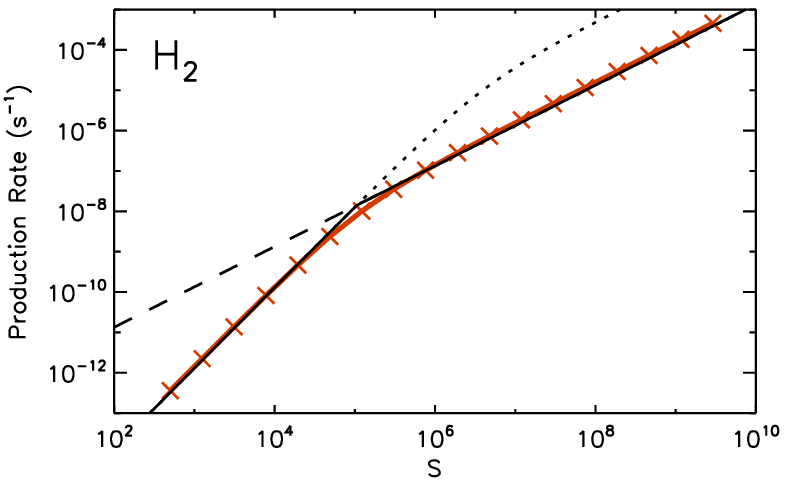}
\caption{Population of H atoms and production rate of H$_2$ on grain surface as a function of $S$, the number of sites per grain, adopting low $R_{acc}(H)$. Red lines are the master-equation results of \cite{barzel2}, with crosses indicating the data points. Solid black lines indicate results of the simple modification scheme of Section 3.1. Dashed lines are the standard rate-equation solutions; dotted lines indicate the solutions using {\em only} equations (11) and (12).}
\label{h1}
\end{figure}

\begin{figure}
\centering
\includegraphics[width=8cm]{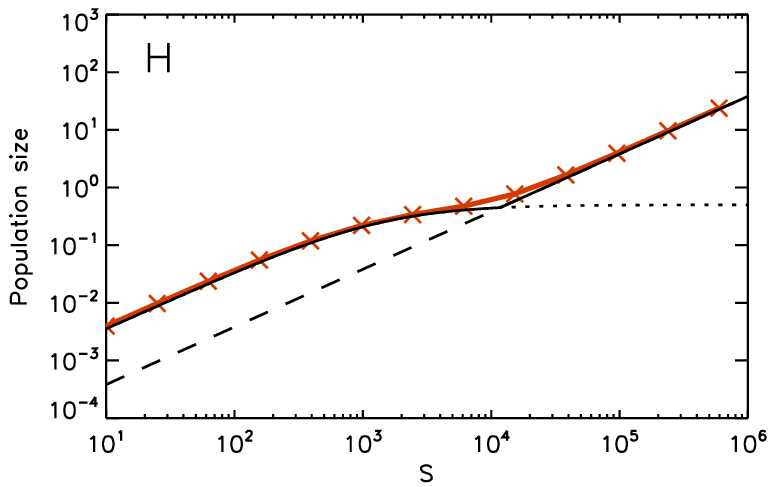}
\includegraphics[width=8cm]{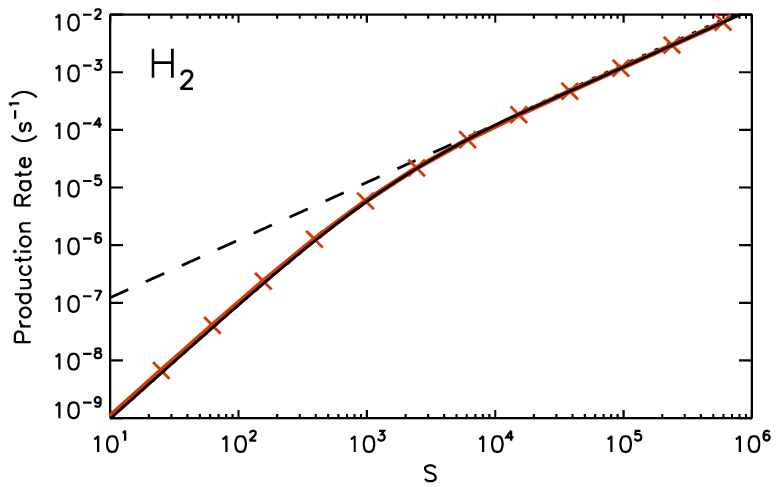}
\caption{Population of H atoms and production rate of H$_2$ on grain surface as a function of $S$, the number of sites per grain, adopting high $R_{acc}(H)$. Red lines are the master-equation results of \cite{barzel2}, with crosses indicating the data points. Solid black lines indicate results of the simple modification scheme of Section 3.1. Dashed lines are the standard rate-equation solutions; dotted lines indicate the solutions using {\em only} equations (11) and (12).}
\label{h2}
\end{figure}

\subsection{Rate-equation inaccuracies}

The expectation value, $\langle N(i) \rangle$, of the grain-surface population of species $i$ may be defined simply as:
\begin{equation}
\langle N(i) \rangle \equiv \sum_{N=0}^{\infty}N \cdot P_{N}(i)
\end{equation} 
where $P_{N}(i)$ is the probability of finding $N$ atoms/molecules of species $i$ on the grain, at any arbitrary moment. The quantity $\langle N(i) \rangle$ may be considered to be the average population of species $i$ over a large ensemble of interstellar dust grains, as suggested by \cite{lipshtat}; but it may also be viewed as the average population on an {\em individual} grain over a statistically long time period. Such an understanding is valid so long as the macroscopic behaviour of the system is in a quasi-steady state in relation to the microscopic processes that occur on grains.

As explained by Biham et al. (2001) and Lipshtat et al. (2004), the inaccuracies of rate equations result from the use of the multiplication products $\langle N(A) \rangle \cdot \langle N(B) \rangle$ and $\frac{1}{2} \langle N(A) \rangle ^{2}$ in the production rates of equations (5) and (6). These quantities are used to represent the average number of unique pairs of reactants present on the grain. They are thus only approximations to the true average number of unique pairs, $\langle N(A) \cdot N(B) \rangle$ and $\frac{1}{2} \langle N(A) \cdot [N(A)-1] \rangle$. The approximations are valid when the surface populations of reactants are large, but {\em may} break down when those values become small. Hence, the equalities,
\begin{equation}
\langle N(A) \rangle \cdot \langle N(B) \rangle = \langle N(A) \cdot N(B) \rangle
\end{equation}
\begin{equation}
\langle N(A) \rangle ^{2} = \langle N(A) \cdot [N(A)-1] \rangle \equiv \langle N^{2}(A) \rangle - \langle N(A) \rangle
\end{equation}
are expressions of the validity of the ``deterministic'' rates shown in equations (5) and (6). However, the condition that $\langle N(A) \rangle$ or $\langle N(B) \rangle$ be small (i.e., of the order of 1 or less) is necessary, {\em but not sufficient}, to invalidate equations (8) and (9). This may clearly be seen from the results of Barzel \& Biham (2007b), in which the deterministic rates are sometimes seen to be accurate even when the populations of all reactants are comfortably less than 1. In the case of equation (8), the equality is valid so long as the probabilities, $P_{N}(A)$ and $P_{N}(B)$, of individual population states, $N(A)$ and $N(B)$, remain uncorrelated. If average populations are small {\em and} reaction rates, given by equation (3), are fast, then these probabilities may become {\em anti-}correlated, invalidating equation (8). In this case, the probability of finding one particle of $A$ and one particle of $B$ on the grain surface at the same time is low, because they quickly react to produce $AB$. Following this argument, the deterministic rates, equations (5) and (6), must represent an absolute upper limit to the production rates, in all regimes.

\begin{table*}
\caption[]{The surface reactions used in each system. Activation energies employed by Stantcheva et al. 2002 are also indicated.}
\label{tab1}
\begin{center}
\begin{tabular}{lcccccc}
\hline
\hline
\noalign{\smallskip}
Reaction  & \multicolumn{3}{c}{Barzel \& Biham (2007b)} && \multicolumn{2}{c}{Stantcheva et al. (2002)}\\
\noalign{\smallskip}
\cline{2-4}\cline{6-7}
\noalign{\smallskip}
          & H$_2$ system & H$_2$O system & CH$_3$OH system && CH$_3$OH system & $E_A$ (K) \\
\noalign{\smallskip}
\hline
\noalign{\smallskip}
H + H       $\rightarrow$ H$_2$       & $\bullet$ & $\bullet$ & $\bullet$ && $\bullet$ &  \\
\noalign{\smallskip}
H + O       $\rightarrow$ OH          &           & $\bullet$ & $\bullet$ && $\bullet$ &  \\
\noalign{\smallskip}
H + OH      $\rightarrow$ H$_2$O      &           & $\bullet$ & $\bullet$ && $\bullet$ &  \\
\noalign{\smallskip}
O + O       $\rightarrow$ O$_2$       &           & $\bullet$ & $\bullet$ && $\bullet$ &  \\
\noalign{\smallskip}
H + CO      $\rightarrow$ HCO         &           &           & $\bullet$ && $\bullet$ & 2000 \\
\noalign{\smallskip}
H + HCO     $\rightarrow$ H$_2$CO     &           &           & $\bullet$ && $\bullet$ &  \\
\noalign{\smallskip} 
H + H$_2$CO $\rightarrow$ CH$_3$O     &           &           & $\bullet$ && $\bullet$ & 2000 \\
\noalign{\smallskip}
H + CH$_3$O $\rightarrow$ CH$_3$OH    &           &           & $\bullet$ && $\bullet$ &  \\
\noalign{\smallskip}
O + CO      $\rightarrow$ CO$_2$      &           &           & $\bullet$ && $\bullet$ & 1000 \\
\noalign{\smallskip}
O + HCO     $\rightarrow$ CO$_2$ + H  &           &           & $\bullet$ && $\bullet$ &  \\
\noalign{\smallskip}
\hline
\end{tabular}
\end{center}
\end{table*}

\section{Basic modifications}

Here, rather than calculate $\langle N(A) \cdot N(B) \rangle$ or $\langle N^{2}(A) \rangle$, the grain-surface production rates are evaluated by an altogether different approach. Firstly, a basic form is constructed for the production rates, with which the standard reaction rates of equations (5) and (6) may be replaced. In order to retain the same matrix-inversion techniques with which the gas-phase equations are integrated, it is essential to express these rates in terms of the averaged populations, $\langle N(i) \rangle$, of each species $i$. 

Consider a system of two reactive species, $A$ and $B$, accreting onto a dust grain at rates $R_{acc}(A)$ and $R_{acc}(B)$, and with evaporation rates $k_{evap}(A)$ and $k_{evap}(B)$. A single reaction, $A + B \rightarrow AB$, is allowed to occur, at a rate $k_{AB}$, as defined in equation (3).

The simplest stochastic case is that in which each reacting species has an average abundance $\langle N(i) \rangle << 1$. Here, the probability of population states $N(i) \geq 2$ is very small, so the probability of finding 1 (or more) of reactant $i$ on the grain at any moment may be approximated as: 
\begin{equation}
P(i) \simeq \langle N(i) \rangle .
\end{equation}

It is assumed that reaction between $A$ and $B$ is fast; i.e., $k_{AB} >> R_{acc}(A), R_{acc}(B), k_{evap}(A), k_{evap}(B)$, so $A$ and $B$ will react as soon as both are present. The reaction rate is therefore unimportant in the calculation of the overall production rate of species $AB$ (thus the rate-equation method fails). It is the accretion rates that determine the production rate; the system is in what is commonly referred to as the ``accretion limit''. The rate of production is simply equal to the rate of accretion of one species times the probability that the other is present, and vice versa. Using equation (10) this gives:
\begin{equation}
R_{mod}(AB)=R_{acc}(B) \cdot \langle N(A) \rangle+R_{acc}(A) \cdot \langle N(B) \rangle .
\end{equation}

If $A$ and $B$ are atoms/molecules of the same species (e.g. atomic H), which may react together, the final result is:
\begin{equation}
R_{mod}(AA)=R_{acc}(A) \cdot \langle N(A) \rangle .
\end{equation}

As long as $k_{AB} >> k_{evap}(A), k_{evap}(B)$, there will be no competition from other processes; hence, expressions (11) and (12) are accurate even if the reaction rate is comparable to either accretion rate. Implicit in the formulation above is the assumption that only the two species crucial to the reaction are present on the grain; no competition with other reactions is considered.  

Equations (11) and (12) may also be easily arrived at by an analytical master-equation approach. Lipshtat et al. (2004) dub this result the ``small grain'' approximation, as low values of $\langle N(A) \rangle$ and $\langle N(B) \rangle$ and fast reaction rates, $k_{AB}$, are achieved as the grain radius asymptotically approaches zero.

Expressions (11) and (12) provide the {\em basic} form of the production rates with which it is proposed to replace the old rate-equation expressions, given the correct conditions. Equations (5) and (6) may be regarded as representative of the extreme situation $\langle N(A) \rangle,\langle N(B) \rangle >> 1$, and equations (11) and (12) as representative of the other extreme, $\langle N(A) \rangle,\langle N(B) \rangle << 1$.

This simple formulation may be compared to that proposed by Caselli et al. (1998) and explored in a number of subsequent papers. They proposed, in their ``corrected'' formulation (Stantcheva et al. 2001), to replace the standard hydrogenation reaction rates, $k_{H,X}$, i.e. the production rate {\em coefficients}, with the faster of either the accretion rate or evaporation rate of atomic hydrogen. In the case where $k_{evap}($H$) > R_{acc}($H$)$ (the so-called ``evaporation limit''), the steady-state abundance of atomic hydrogen may be evaluated as $\langle N($H$) \rangle \simeq R_{acc}($H$) / k_{evap}($H$)$, yielding a modified production rate of $R_{acc}($H$) \cdot \langle N($X$) \rangle$. For the reaction H + H $\rightarrow$ H$_2$, this form is equal to equation (12). For other reactions (with X$\neq$H), $\langle N(H) \rangle$ is expected to be very small, so this same form is also approximately equal to equation (11). Hence, the modifications for hydrogenation reactions proposed by Caselli et al. (1998) and Stantcheva et al. (2001) for the ``evaporation limit'' case may be seen to be a special case of equations (11) and (12). Those modifications remain accurate so long as the ``small grain'' limit is maintained, and evaporation is still the dominant removal mechanism for atomic hydrogen.

\begin{table*}
\caption[]{Binding energies, diffusion barriers, fluxes, and dust temperatures employed by Barzel \& Biham (2007a,b)}
\label{tab2}
\begin{center}
\begin{tabular}{lcccccccccccc}
\hline
\hline
\noalign{\smallskip}
Species  & \multicolumn{4}{c}{H$_2$ system} && \multicolumn{3}{c}{H$_2$O system} && \multicolumn{3}{c}{CH$_3$OH system}\\
\noalign{\smallskip}
         & \multicolumn{4}{c}{($T_{d} = 10$ K)} && \multicolumn{3}{c}{($T_{d} = 15$ K)} && \multicolumn{3}{c}{($T_{d} = 15$ K)}\\
\noalign{\smallskip}
\cline{2-5}\cline{7-9}\cline{11-13}
\noalign{\smallskip}
         & $E_D$ (K) & $E_b$ (K) &  \multicolumn{2}{c}{$R_{acc}/S$ (s$^{-1}$)} && $E_D$ (K) & $E_b$ (K) & $R_{acc}/S$ (s$^{-1}$) && $E_D$ (K) & $E_b$ (K) & $R_{acc}/S$ (s$^{-1}$) \\
\noalign{\smallskip}
\hline
\noalign{\smallskip}
H        & 371 & 255 & $1.00 \times 10^{-11}$ & $2.75 \times 10^{-8}$ && 603 & 511 & $5.0 \times 10^{-10}$ && 603 & 511 & $5.0 \times 10^{-10}$ \\
\noalign{\smallskip}
O        &  &        & &                                              && 627 & 545 & $1.0 \times 10^{-11}$ && 627 & 545 & $1.0 \times 10^{-11}$ \\
\noalign{\smallskip}
OH       &  &        & &                                              && 627 & 545 &      --               && 627 & 545 & --  \\
\noalign{\smallskip}
CO       &  &        & &                                              &&  &        &                       && 638 & 580 & $1.0 \times 10^{-10}$ \\
\noalign{\smallskip}
HCO      &  &        & &                                              &&  &        &                       && 673 & 603 & --  \\
\noalign{\smallskip} 
H$_2$CO  &  &        & &                                              &&  &        &                       && 685 & 615 & --  \\
\noalign{\smallskip}
CH$_3$O  &  &        & &                                              &&  &        &                       && 719 & 638 & --  \\
\noalign{\smallskip}
\hline
\end{tabular}
\end{center}
\begin{list}{}{}
\item{$S$ equals the number of binding sites on the grain}
\end{list}
\end{table*}

\subsection{Rate replacement and restrictions}

To determine whether the production rates of equations (5) and (6) should be replaced with equations (11) and (12), the validity of equations (8) and (9) must be assessed. If $\langle N(A) \rangle >> 1$ and $\langle N(B) \rangle >> 1$ then the equalities, and the deterministic rates, are valid. In the case that, for example, $\langle N(A) \rangle >> 1$ but $\langle N(B) \rangle << 1$, equation (8) is still valid, and rate equations may be used. (An assumption to this effect was made by Stantcheva et al. 2002, in treating species such as CO using deterministic rates). In such cases, reactions involving species $B$ would have only a small effect on the population state of species $A$, making any \linebreak (anti-)correlation very weak. In the case where both $\langle N(A) \rangle << 1$ and $\langle N(B) \rangle << 1$ the rate substitutions should be made.

To apply these conditions, species with $\langle N(i) \rangle \geq 1$ are deemed to be in the $\langle N(i) \rangle >> 1$ regime, whilst species with $\langle N(i) \rangle < 1$ are deemed to be in the $\langle N(i) \rangle << 1$ regime. Modifications are therefore made only when $\langle N(A) \rangle,\langle N(B)\rangle < 1$. This is, of course, a gross simplification; however, the tests to follow demonstrate that any resultant inaccuracies are small.

As a further restriction on the new rates, it is asserted that under no circumstances may the modified production rate exceed the standard rate-equation value, following the argument of Section 2.1. It is therefore required that:
\begin{equation}
R_{mod}(AB) \leq k_{AB} \cdot \langle N(A) \rangle \cdot \langle N(B) \rangle .
\end{equation}

Thus, if a modified production rate exceeds the deterministic rate, the deterministic rate is used.

\subsection{The hydrogen system}

This basic formulation is tested against the simplest of grain-surface systems, in which atomic hydrogen is the only reactive species. Barzel \& Biham (2007b) investigated this system for various grain sizes, at a temperature of 10 K. They conducted rate-equation, master-equation, and moment-equation simulations, run to steady state in the population of atomic hydrogen. Calculations were made using either a high or a low flux of accreting H-atoms; $R_{acc}(H)=2.75 \times 10^{-8} S$ s$^{-1}$ or $R_{acc}(H)=1 \times 10^{-11} S$ s$^{-1}$, where $S$ is the number of surface binding sites. Details of this system are given in Tables 1 \& 2. \cite{barzel2} detail all other relevant values.

Solving for H populations and H$_2$ production rates requires the solution of the equation:
\begin{eqnarray*}
\frac{d \langle N(\mathrm{H}) \rangle}{dt} = R_{acc}(\mathrm{H}) - k_{evap}(\mathrm{H}) \cdot \langle N(\mathrm{H}) \rangle - R_{prod}(\mathrm{H}_{2})=0 
\end{eqnarray*}
where $R_{prod}($H$_{2})$ is defined according to the stipulations of Section 3.1. Obtaining an analytical solution is trivial. Figures 1 and 2 show grain-surface atomic hydrogen populations and H$_2$ production rates calculated in this way for various values of $S$, using low and high H-fluxes, respectively. Solid black lines show the results when modified rates are employed according to the stipulations of the modification scheme of Section 3.1. Dashed lines represent standard rate-equation results; dotted lines represent results obtained {\em purely} from equations (11) and (12). Indicated in red are the master-equation results of \cite{barzel2}, with crosses marking the individual data points. The master-equation results may be regarded as a true representation of the system.

Rates and populations calculated using standard rate equations rise linearly with increasing $S$. At large grain sizes these are the exact solutions, but become inaccurate for small grains. Similarly, populations calculated purely with equations (11) and (12) may diverge from the rate equations for large grain sizes. 

The simple modification scheme demonstrates near-perfect agreement with the master-equation populations and production rates calculated by Barzel \& Biham, at both the large- and small-grain extremes of the system, for each H-flux value. Approaching the stochastic--deterministic threshold, where deterministic rates become accurate, results vary marginally from the exact master-equation values; the match is otherwise perfect.

\section{Continuous modification schemes}

Whilst modification using equations (10) --  (13), along with the stipulations of Section 3.1, is accurate in the case of simple, highly prescribed systems, the purpose of this work is to fashion a modification scheme that is more universally applicable to large gas-grain chemical networks. Such a scheme must provide rate continuity over the stochastic--deterministic threshold at $\langle N(i) \rangle = 1$. Any functional form must also be integrable using standard differential equation-solving techniques (i.e. the Gear algorithm). Below, further modifications to, and restrictions on, the basic equations (11) and (12) are formulated.

\subsection{Threshold continuity}
 
The switch-over between the modified rate and the standard rate may produce a discontinuity in the rates at $\langle N(i) \rangle =1$. Dependent on the relative rates of all the processes involved, this may present an impediment to accurate calculations. Therefore, a simple empirical function, $f$, is introduced to make the transition smoother whilst preserving a fast switch-over. Under this scheme, production rates are {\em always} modified according to:
\begin{equation}
R_{tot} = f_{AB} \cdot R_{mod}(AB) + (1- f_{AB}) \cdot k_{AB} \cdot \langle N(A) \rangle \cdot \langle N(B) \rangle
\end{equation}
where:
\begin{eqnarray}
\langle N(A) \rangle<1, \langle N(B) \rangle<1:& & f_{AB}=1 \nonumber \\
\langle N(A) \rangle>1, \langle N(B) \rangle<1:& & f_{AB}=1/ \langle N(A) \rangle \nonumber \\
\langle N(A) \rangle<1, \langle N(B) \rangle>1:& & f_{AB}=1/ \langle N(B) \rangle \nonumber \\
\langle N(A) \rangle>1, \langle N(B) \rangle>1:& & f_{AB}=1/[\langle N(A) \rangle \cdot \langle N(B) \rangle] .
\end{eqnarray}
This has the effect that when $\langle N(A) \rangle, \langle N(B) \rangle < 1$, the rate is always ``stochastic'', whilst quickly tending towards the deterministic rate as either $\langle N(A) \rangle$ or $\langle N(B) \rangle$ rises above unity. For example, if species $A$ and $B$ both attain abundances of 10 atoms/molecules per grain, the deterministic contribution to the total production rate will be 99\% of the normal deterministic rate. Expressions (14) \& (15) therefore allow the modified rate contribution to replace a fraction of the total deterministic rate that corresponds to the reaction of 1 atom/molecule of species $A$ with 1 atom/molecule of species $B$. Equation (13) is also applied, meaning that if $R_{mod}(AB) > k_{AB} \cdot \langle N(A) \rangle \cdot \langle N(B) \rangle$, the total rate is equal to the unmodified deterministic rate.

\begin{figure}
\centering
\includegraphics[width=8cm]{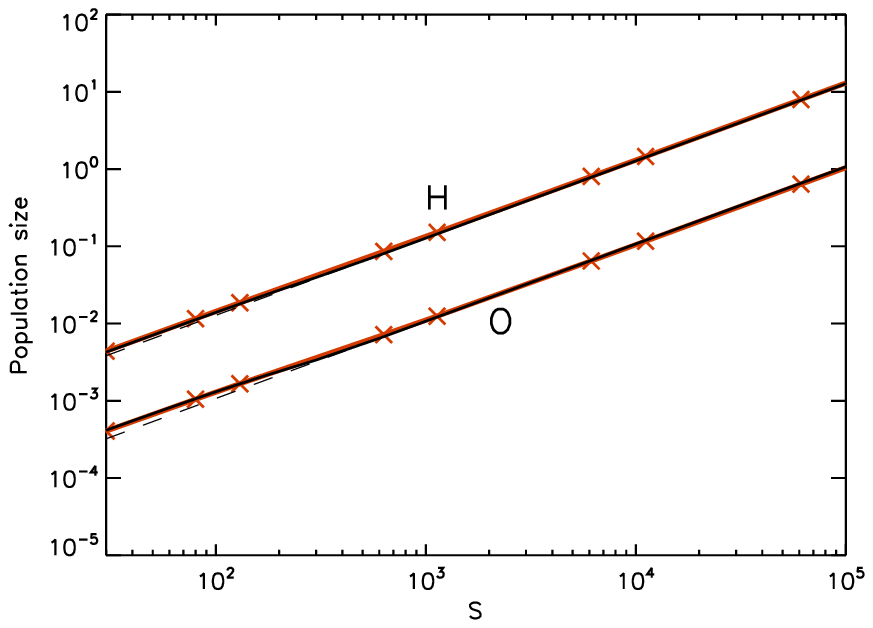}
\includegraphics[width=8cm]{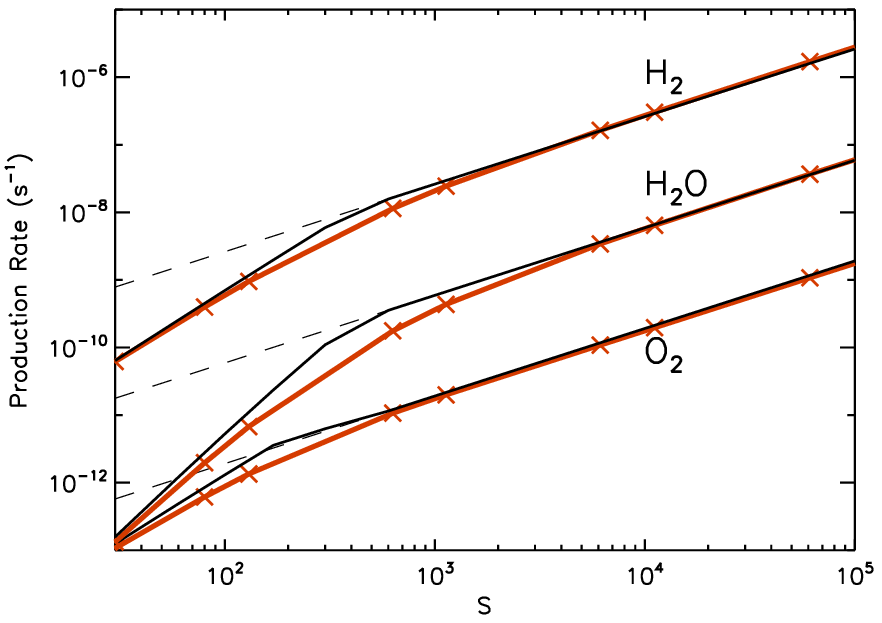}
\caption{Populations and production rates for the H$_2$O system, assuming no reaction competition. Red lines are the master-equation results of \cite{barzel2}, with crosses indicating the data points. Solid black lines are the results of the modification scheme of Section 4.1. Dashed lines indicate the standard rate-equation results.}
\end{figure}

\begin{figure}
\centering
\includegraphics[width=8cm]{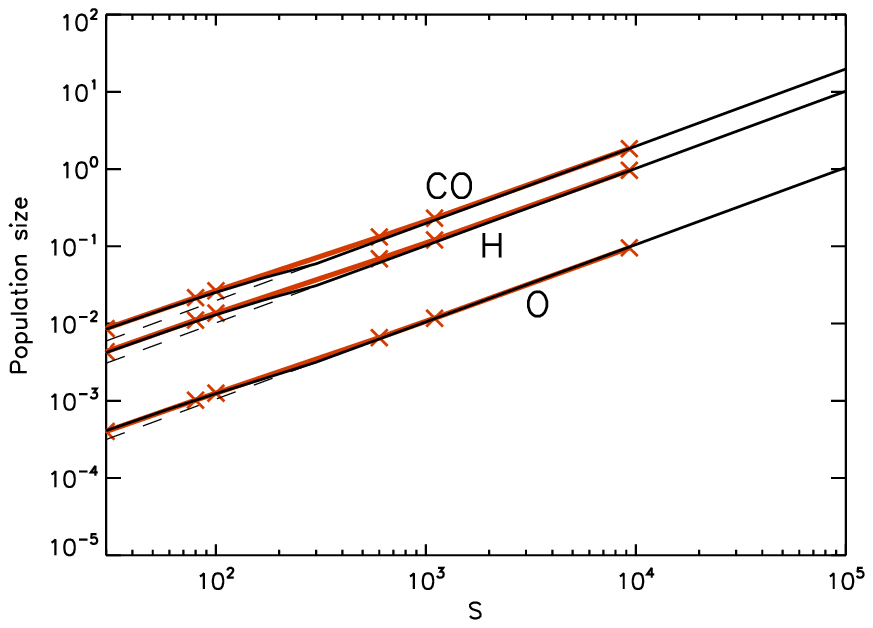}
\includegraphics[width=8cm]{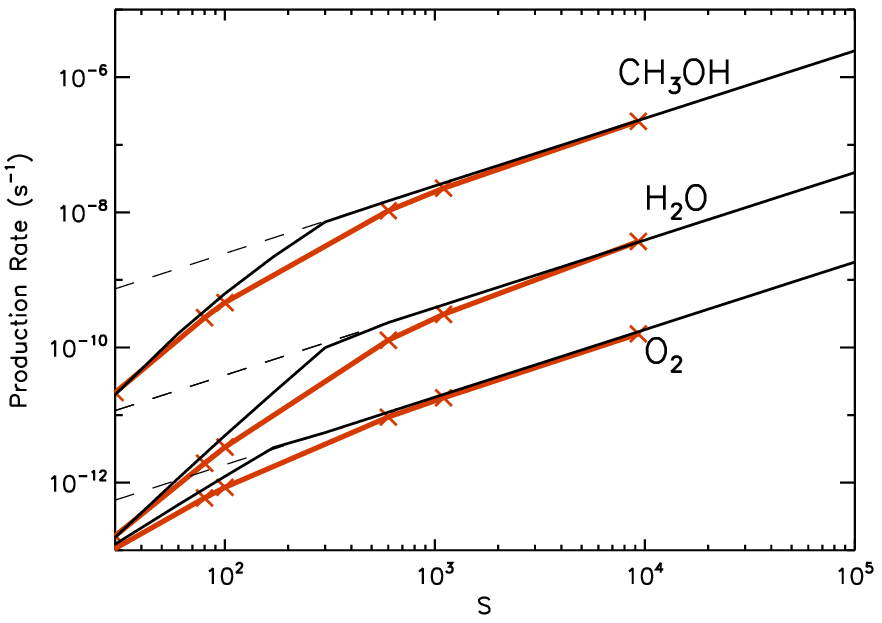}
\caption{Populations and production rates for the CH$_3$OH system, assuming no reaction competition. Red lines are the master-equation results of \cite{barzel2}, with crosses indicating the data points. Solid black lines are the results of the modification scheme of Section 4.1. Dashed lines indicate the standard rate-equation results.}
\end{figure}

\subsection{The water system and the methanol system}

The continuous rate-modification scheme outlined above is used to examine the water and methanol systems, at 15 K, as investigated by \cite{barzel2}. The former scheme includes reactions between surface species H, O, and OH, resulting in H$_2$, O$_2$ and H$_2$O production. The latter scheme also includes  reactions with CO, HCO, H$_2$CO, and CH$_3$O, leading to the production of methanol, CH$_3$OH, and carbon dioxide, CO$_2$. The reactions, fluxes, and binding energies are indicated in Tables 1 \& 2. It should be noted that the methanol system investigated by Barzel \& Biham does not include any activation-energy barriers. In fact, activation energies substantially complicate the behaviour of the methanol system; see Section 6.

\begin{figure*}
\centering
\includegraphics[width=8cm]{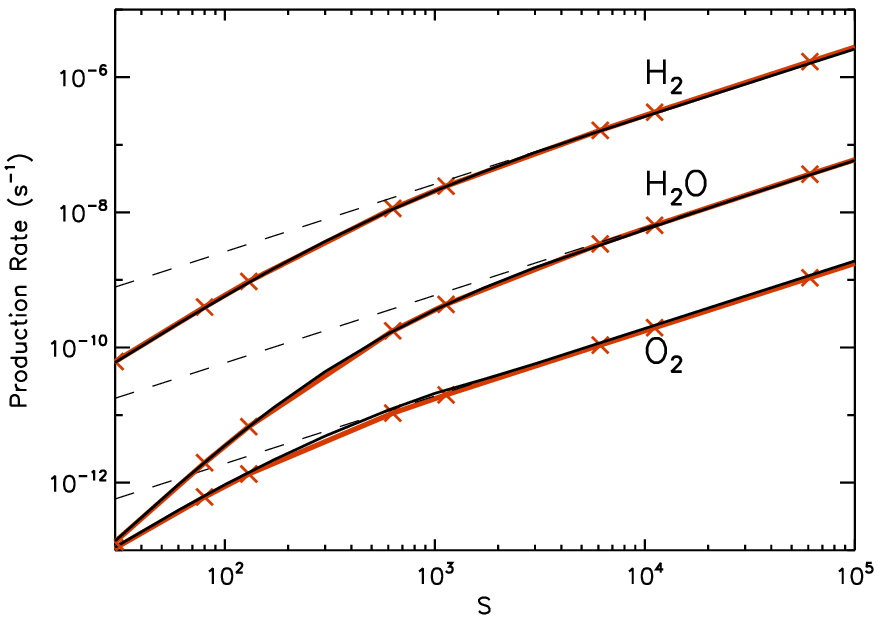}
\includegraphics[width=8cm]{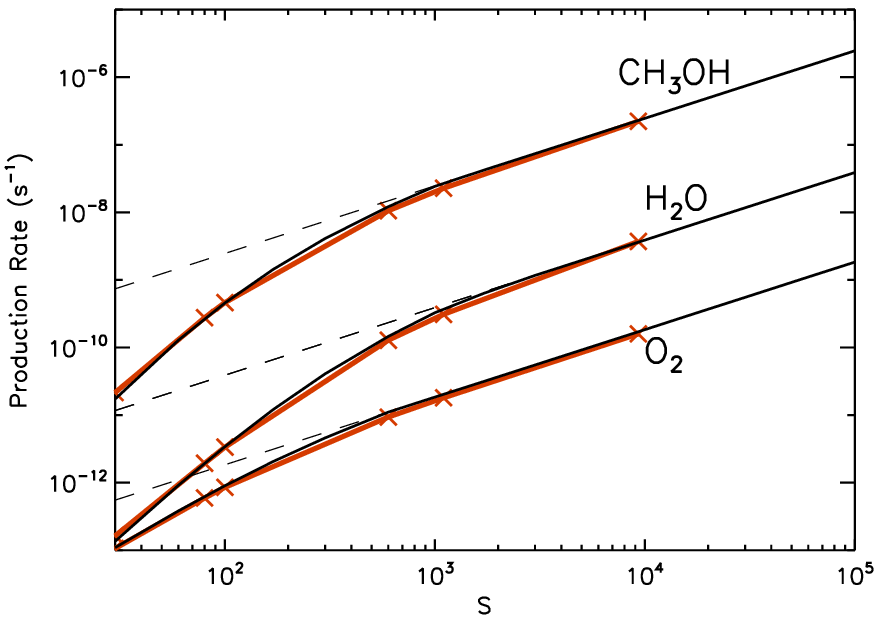}
\caption{Production rates for the H$_2$O system (left panel) and CH$_3$OH system (right panel), using the simple competition scheme of Section 5. Red lines are the master-equation results of \cite{barzel2}, with crosses indicating the data points. Solid black lines show the results of the modification scheme of Section 5. Dashed lines indicate the standard rate-equation results. The apparent small deviations between master-equation and modified-rate methods derive largely from differences in $S$-sampling.}
\end{figure*}

Figures 3 \& 4 show population sizes and production rates of key species, for the water and methanol systems, respectively. Populations obtained from the new method (Section 4.1) are very well matched to the master-equation results of Barzel \& Biham. Production rates are accurate at high and very low values of $S$, but values near the stochastic--deterministic threshold are more obviously inaccurate, whilst obeying the correct trend. However, the new method is a very good first approximation.

What are the underlying physical reasons for the disagreements? The production rates obtained from the new method rise to the rate equation values at lower values of $S$ than the master-equation results. This occurs before any reactants approach a population of 1, as the rate limit of equation (13) is reached before this. Hence, the empirical function that is used to switch over at the $\langle N(i) \rangle = 1$ threshold is not the cause.

In fact, the modified rates are too fast because competition between surface processes has not been considered. At low values of $S$, reactions are extremely fast, due to the fast reaction rates of all reactants; see equation (3). As $S$ increases, reaction becomes slower, and the possibility arises that one or other reactant may evaporate before the two meet in the same binding site and react. For hydrogen atoms in the water and methanol systems, using the binding energies shown in Table 2, $k_{HX}=k_{evap}(H)$ when $S \simeq 460$ sites per grain. Discrepancies in hydrogen-related production rates are greatest close to this point, as may be seen in figures 3 \& 4. To accurately evaluate the production rates, competition processes must be considered.

\section{Evaporation--Reaction Competition}

The production rates expressed in equations (11) \& (12) are simply the rates at which one or other reacting species accretes onto the grain when the other is present on the surface. This assumes 100\% efficiency in the reaction that follows the accretion event. 

To take account of competition in the modified rates, an efficiency factor, $C_{AB}$, is employed, such that:
\begin{equation}
R_{mod}(AB)=C_{AB} \cdot (R_{acc}(B) \cdot \langle N(A) \rangle+R_{acc}(A) \cdot \langle N(B) \rangle)
\end{equation}
\begin{equation}
C_{AB}=k_{AB}/[k_{AB}+k_{evap}(A)+k_{evap}(B)] .
\end{equation}
The efficiency, $C_{AB}$, represents the probability that, of three possible processes -- reaction of A and B, evaporation of A, and evaporation of B -- it is reaction that occurs. This assumes only two reacting atoms/molecules to be present at any one time; an assumption made by many other authors in considering such systems (e.g. Allen \& Robinson 1977, Green et al. 2001, Barzel \& Biham 2007). Hence, reactions with other (stochastic) species may be ignored, for the purposes of competition.

It should be noted that, when valid, the rate equations naturally take account of any form of competition that may arise. 

Applying the efficiency factor to $R_{mod}(AB)$ for all reactions in the H$_2$O and CH$_3$OH systems gives the results shown in figure 5. Production rates now show precisely the behaviour expected, and provide an excellent match to the master-equation results, within computational accuracy. (Population sizes do not vary noticeably from the values shown in figures 3 \& 4, and are hence omitted). Some small discrepancies seem to appear in the production rates, particularly for the CH$_3$OH system; however, these derive largely from greater $S$-sampling in the new modified-rate models than in the master-equation models.

\begin{table*}
\caption[]{Binding energies, diffusion barriers, and fluxes employed by Stantcheva et al. (2002)}
\label{tab3}
\begin{center}
\begin{tabular}{lcccccc}
\hline
\hline
\noalign{\smallskip}
\noalign{\smallskip}
Species  & $E_D$ (K) & $E_b$ (K) && \multicolumn{3}{c}{$R_{acc}$ (s$^{-1}$)} \\
\noalign{\smallskip}
\cline{5-7}
\noalign{\smallskip}
         &           &           && Low $n_H$ & Intermediate $n_H$ & High $n_H$ \\
\noalign{\smallskip}
\hline
\noalign{\smallskip}
H         & 350  & 100 && $1.67 \times 10^{-5}$ & $1.67 \times 10^{-5}$ & $1.60 \times 10^{-5}$ \\
\noalign{\smallskip}
O         & 800  & 240 && $3.26 \times 10^{-7}$ & $2.72 \times 10^{-6}$ & $2.53 \times 10^{-5}$ \\
\noalign{\smallskip}
OH        & 1260 & 378 && -- & -- & -- \\
\noalign{\smallskip}
H$_2$     & 450  & 135 && -- & -- & -- \\
\noalign{\smallskip}
O$_2$     & 1210 & 363 && -- & -- & -- \\
\noalign{\smallskip}
H$_2$O    & 1860 & 558 && -- & -- & -- \\
\noalign{\smallskip}
CO        & 1210 & 363 && $2.05 \times 10^{-7}$ & $2.05 \times 10^{-6}$ & $2.05 \times 10^{-5}$ \\
\noalign{\smallskip}
HCO       & 1510 & 453 && -- & -- & -- \\
\noalign{\smallskip}
H$_2$CO   & 1760 & 528 && -- & -- & -- \\
\noalign{\smallskip}
CH$_3$O   & 2170 & 651 && -- & -- & -- \\
\noalign{\smallskip}
CH$_3$OH  & 2060 & 618 && -- & -- & -- \\
\noalign{\smallskip}
CO$_2$    & 2500 & 750 && -- & -- & -- \\
\noalign{\smallskip}
\hline
\end{tabular}
\end{center}
\end{table*}

\subsection{Anterior Competition and Formation Rates}

In the simple hydrogen system, the only source of hydrogen atoms on the grains is direct accretion from the gas phase. However, in the methanol system, hydrogen atoms may be produced by chemical reaction. Furthermore, important radicals like HCO are formed solely on the grains, with no contribution from accretion. Thus, for the reaction H + HCO $\rightarrow$ H$_2$CO, equation (12) contains the single term $R_{acc}($H$) \cdot \langle N($HCO$) \rangle$. But, should HCO formation by reaction also be thought of as an ``accretion'' process, for the purposes of equation (12), leading to a term $R_{form}($HCO$) \cdot \langle N($H$) \rangle$?

The inclusion of the term $R_{form}($HCO$) \cdot \langle N($H$) \rangle$ would assume an H-atom to be present on the grain whilst HCO itself were formed. However, since HCO is formed by the reaction H + CO $\rightarrow$ HCO, this would mean the presence of two H-atoms on the grain at the same time, prior to HCO formation. Rather than forming HCO, the most likely outcome is that the two hydrogen atoms should react together instead. To treat this situation accurately, competition between reactions prior to the reaction of interest (i.e. H + HCO $\rightarrow$ H$_2$CO) should be taken into account. Such competition is labelled here as ``anterior competition''. 

In fact, if the two-particle approximation (e.g. Section 5) is valid, the problem of anterior competition may be almost entirely avoided. The two-particle assumption precludes the possibility that one reactant may be formed on the grains in the presence of the other. This allows the formation processes resulting from {\em stochastic} reactions to be omitted from equations (11) and (12), or equation (16), and anterior competition ignored.

However, in the case where, for example, CO is present in great abundance on the grains, the reactions of CO would be deterministic in nature, and CO should be considered present at all times. Thus, the deterministic formation of HCO by surface reactions should be included in the ``accretion'' rates utilised in equations (11) and (12), or equation (16). 

In the models that follow, $R_{acc}$ is substituted with $R_{form}$.
\begin{equation}
R_{form}(A) = R_{acc}(A) + \sum_{all \ ij \ pairs} (1- f_{ij}) \cdot k_{ij} \cdot \langle N(i) \rangle \cdot \langle N(j) \rangle 
\end{equation}
for all reaction pairs $ij$ that form species $A$, where $k_{ij}$ is the reaction rate, and $f_{ij}$ is defined in Section 4.1. As the deterministic part of the production rate in equation (14) is unaffected by the stochastic part, there is no need to iterate the calculations.

For consistency, equation (17) should include, in the denominator, terms for the reaction of particles $A$ and $B$ with other species, such as CO, that can build up significant abundances on the grains. If CO, or any other species, has $\langle N(i) \rangle >> 1$ then there is a vanishing probability of {\em not} finding such a reactant on the grains when two other particles are also present. Reactions with that species must be allowed to compete with the reaction between $A$ and $B$, if such reactions exist. 

In this model, $\langle N(i) \rangle =1$ is used as the stochastic--deterministic threshold. If $\langle N(i) \rangle > 1$ then terms of the form $\left( k_{ij} \cdot \left[ \langle N(i) \rangle -1 \right] \right)$ should be inserted into equation (17) for the reactions of any species $j$ with which species $i$ may react. This means that only the deterministic part of the reaction rate is involved in the competition. This treatment is in keeping with the formalism of equations (14) and (15) and Section 4.1.

In fact, this final precaution actually has no significant effect on the systems modelled in this paper; the activation energy barriers used later, in Section 6, are too high to make reaction with CO or H$_2$CO sufficiently competitive to affect the reactions of atomic hydrogen or oxygen. However, for adoption in a generalised system, this eventuality is easily treated.

\section{Systems with activation energies \label{act-sec}}

Activation energy barriers are typically assumed to mediate key reactions in the methanol system. However, these barriers were not considered in the study of \cite{barzel2}. In order to test the new modified-rate method against a system with activation energies, the methanol system of \cite{stant2} is employed; see Tables 1 and 3. 

\cite{stant2} used a hybrid master-equation/rate-equation method to reproduce Monte Carlo results. The former approach assumed that the very reactive species H, O, OH, HCO and CH$_3$O behave stochastically, but that species with reactions requiring activation energies (CO and H$_2$CO), which typically achieve large populations, could be treated using the standard rate equations. The cut-offs for stochastic species were set to values 1, 2, or 3. Three basic scenarios were considered, characterised by the accretion fluxes of H, O, and CO atoms/molecules, mimicking conditions in interstellar clouds of ``low'', ``intermediate'', and ``high'' density. The models were run to 1000 yr of evolution, assuming $10^6$ binding sites per grain and a grain temperature of 10 K. Importantly, Stantcheva et al. allowed diffusion of hydrogen atoms around the grain surface to occur via quantum tunnelling between binding sites, rendering the rates of hydrogen reaction substantially faster than would be the case assuming only thermal hopping. \cite{katz} suggested that such quantum-tunnelling effects do not govern surface diffusion rates. However, the purpose of the comparison is, in this case, to test the method against a reliable standard, rather than to provide an accurate reproduction of grain-surface chemistry in interstellar clouds. More rigorous application of the new modification methods to interstellar cloud conditions will follow in future.

Of the two methods utilised by Stantcheva et al., the Monte Carlo results should be considered the most reliable, although the differences are typically not large. However, the Monte Carlo simulations use only integer values, so it is not possible to obtain accurate estimates of average population sizes when those values are close to unity. Comparison with species of very low abundance relies on the results of the  master-equation/rate-equation hybrid method alone. For convenience, this method will from now on be refered to simply as the ``master equation'' method, with the understanding that it is in fact a hybrid scheme.

\begin{table*}
\caption[]{Modified rate results for low density, using input values from Stantcheva et al. (2002). Values in boldface show agreement within 10\% of the Monte Carlo or master-equation values.}
\label{tab4}
\begin{center}
\begin{tabular}{lrccccccccccc}
\hline
\hline
\noalign{\smallskip}
Species   & \multicolumn{2}{c}{Stantcheva et al.} && Rate Eq. && \multicolumn{7}{c}{Modified Rates} \\
\noalign{\smallskip}
\cline{2-3}\cline{5-5}\cline{7-13}
\noalign{\smallskip}
          & Monte Carlo  & Master Eq. 22211 &&   && Method A    && Method B   && Method C          && Method D    \\
\noalign{\smallskip}
\cline{7-7}\cline{9-9} \cline{11-11}\cline{13-13}
\noalign{\smallskip}
          &              &          &&                   && ``Simple''         && ``Full''           && Method B +         && Method C +  \\
          &              &          &&                   && Competition        && Competition        && Poisson Prob.      && H-accretion \\
\noalign{\smallskip}    
\hline
\noalign{\smallskip}
H         & 1            & 7.96(-3) &&  1.22(-5)         && {\bf 7.97(-3)}     && {\bf 7.97(-3)}     && {\bf 7.97(-3)}     && {\bf 7.96(-3)} \\
\noalign{\smallskip}
O         & 0            & 1.90(-2) &&  5.22(-7)         && {\bf 1.85(-2)}     && {\bf 1.85(-2)}     && {\bf 1.86(-2)}     && {\bf 1.88(-2)} \\
\noalign{\smallskip}
OH        & 0            & 1.93(-2) &&  5.22(-7)         && {\bf 1.86(-2)}     && {\bf 1.86(-2)}     && {\bf 1.88(-2)}     && {\bf 1.90(-2)} \\
\noalign{\smallskip}
H$_2$     & 2            & 1.94(+0) &&  1.11(+2)         && {\bf 1.94(+0)}     && {\bf 1.94(+0)}     && {\bf 1.93(+0)}     && {\bf 1.93(+0)} \\
\noalign{\smallskip}
O$_2$     & 1.70(+2)     & 1.62(+2) &&  3.65(-7)         && 1.89(+2)           && 1.89(+2)           && 1.89(+2)           && {\bf 1.60(+2)} \\
\noalign{\smallskip}
H$_2$O    & 9.90(+3)     & 9.86(+3) &&  {\bf 1.03(+4)}   && {\bf 9.76(+3)}     && {\bf 9.76(+3)}     && {\bf 9.76(+3)}     && {\bf 9.85(+3)} \\
\noalign{\smallskip}
CO        & 0            & 2.81(-2) &&  1.85(+1)         && 3.76(-2)           && {\bf 2.82(-2)}     && {\bf 2.82(-2)}     && {\bf 2.82(-2)} \\
\noalign{\smallskip}
HCO       & 0            & 1.22(-2) &&  3.29(-7)         && {\bf 1.21(-2)}     && {\bf 1.21(-2)}     && {\bf 1.22(-2)}     && {\bf 1.22(-2)} \\
\noalign{\smallskip}
H$_2$CO   & 0            & 2.83(-2) &&  1.89(+1)         && 3.74(-2)           && {\bf 2.82(-2)}     && {\bf 2.82(-2)}     && {\bf 2.84(-2)} \\
\noalign{\smallskip}
H$_3$CO   & 0            & 1.23(-2) &&  3.29(-7)         && {\bf 1.21(-2)}     && {\bf 1.21(-2)}     && {\bf 1.22(-2)}     && {\bf 1.22(-2)} \\
\noalign{\smallskip}
CH$_3$OH  & 6.40(+3)     & 6.39(+3) &&  {\bf 6.42(+3)}   && {\bf 6.34(+3)}     && {\bf 6.34(+3)}     && {\bf 6.34(+3)}     && {\bf 6.37(+3)} \\
\noalign{\smallskip}
CO$_2$    & 90           & 8.95(+1) &&  2.29(-7)         && 1.24(+2)           && 1.24(+2)           && 1.24(+2)           && {\bf 8.97(+1)} \\
\noalign{\smallskip}
\hline
\end{tabular}
\end{center}
\end{table*}

Table 4 shows two sets of results from Stantcheva et al.; the column headed ``Master Equation 22211'' indicates results from their model using cut-offs for stochastic species H, O, OH, HCO, CH$_3$O of 2, 2, 2, 1, 1, respectively. Columns in tables 5 and 6 are labelled similarly. Stantcheva et al. also ran models with lower cut-offs; the results of the models with the highest cut-offs are quoted, on the assumption that these are the most accurate.

Tables 4 -- 6 also show rate-equation solutions for each regime. Whilst a few rate-equation results are within an order of magnitude of the exact value, many populations vary wildly from the Monte Carlo and master-equation results, including important species such as formaldehyde, methanol and CO$_2$. In fact, the rate-equation results shown in Tables 4 -- 6 differ from the values quoted by \cite{stant2}, which actually correspond to activation energies of 2500 K, for the H + CO and H + H$_2$CO reactions, rather than the stated 2000 K. Nevertheless, it is clear that the rate equations produce a very poor match to the exact solutions of each system.

In Table 4, the column headed ``Method A'' shows low-density results using the simple competition scheme of Section 5. Aside from O$_2$, CO, H$_2$CO and CO$_2$, all species show an excellent match to the Monte Carlo and/or master equation results. The worst, CO$_2$, is only a factor of $\sim$1.4 greater than the master-equation value; a great improvement over the rate-equations.

Table 5 shows results for the intermediate density case. Here, the simple competition scheme produces a good match with all species except CO$_2$, which is inaccurate by a factor similar to that of the low-density case. The quality of the match for all other species is not in general as good as for the low-density case, but is still quite acceptable.

Table 6 shows results for the high density case. The results of the simple competition scheme are generally within an order of magnitude of the Monte Carlo and/or master-equation results, and some are rather closer; O$_2$ is very accurate.

In all regimes, method A provides at least a reasonable match to the exact results, and great improvements over the rate-equation method. However, there are still discrepancies that cannot be explained merely as computational inaccuracies between the different methods. Further refinements to the treatment of surface processes are required to match the exact results.

\begin{table*}
\caption[]{Modified rate results for intermediate density, using input values from Stantcheva et al. (2002). Values in boldface show agreement within 10\% of the Monte Carlo or master-equation values.}
\label{tab5}
\begin{center}
\begin{tabular}{lrccccccccccc}
\hline
\hline
  \noalign{\smallskip}
Species   & \multicolumn{2}{c}{Stantcheva et al.} && Rate Eq. && \multicolumn{7}{c}{Modified Rates} \\
\noalign{\smallskip}
\cline{2-3}\cline{5-5}\cline{7-13}
\noalign{\smallskip}
          & Monte Carlo  & Master Eq. 22211 &&   && Method A    && Method B   && Method C          && Method D    \\
\noalign{\smallskip}
\cline{7-7}\cline{9-9} \cline{11-11}\cline{13-13}
\noalign{\smallskip}
          &              &          &&             && ``Simple''         && ``Full''           && Method B +         && Method C +  \\
          &              &          &&             && Competition        && Competition        && Poisson Prob.      && H-accretion \\
\noalign{\smallskip}    
\hline
\noalign{\smallskip}
H         & 1            & 2.88(-3)  &&  5.41(-6)        && {\bf 3.10(-3)}     && {\bf 3.10(-3)}     && {\bf 3.10(-3)}     && {\bf 2.82(-3)} \\
\noalign{\smallskip}
O         & 0            & 1.36(-1)  &&  9.78(-6)        && 1.10(-1)           && 1.10(-1)           && 1.16(-1)           && {\bf 1.25(-1)} \\
\noalign{\smallskip}
OH        & 0            & 1.35(-1)  &&  9.78(-6)        && 1.10(-1)           && 1.10(-1)           && 1.17(-1)           && {\bf 1.25(-1)} \\
\noalign{\smallskip}
H$_2$     & 0, 1         & 7.01(-1)  &&  2.20(+1)        && {\bf 7.55(-1)}     && {\bf 7.55(-1)}     && {\bf 7.54(-1)}     && {\bf 6.85(-1)} \\
\noalign{\smallskip}
O$_2$     & 9.40(+3)     & 9.03(+3)  &&  1.26(-4)        && {\bf 9.36(+3)}     && {\bf 9.36(+3)}     && {\bf 9.36(+3)}     && {\bf 8.44(+3)} \\
\noalign{\smallskip}
H$_2$O    & 6.02(+4)     & 5.93(+4)  &&  8.55(+4)        && {\bf 5.77(+4)}     && {\bf 5.77(+4)}     && {\bf 5.77(+4)}     && {\bf 6.18(+4)} \\
\noalign{\smallskip}
CO        & 1            & 7.76(-1)  &&  4.15(+2)        && {\bf 7.24(-1)}     && {\bf 7.24(-1)}     && {\bf 7.24(-1)}     && {\bf 7.97(-1)} \\
\noalign{\smallskip}
HCO       & 0            & 1.14(-1)  &&  7.39(-6)        && {\bf 1.06(-1)}     && {\bf 1.06(-1)}     && {\bf 1.12(-1)}     && {\bf 1.17(-1)} \\
\noalign{\smallskip}
H$_2$CO   & 1            & 7.11(-1)  &&  4.24(+2)        && 6.35(-1)           && 6.35(-1)           && 6.35(-1)           && {\bf 7.27(-1)} \\
\noalign{\smallskip}
H$_3$CO   & 0            & 1.22(-1)  &&  7.39(-6)        && 1.06(-1)           && 1.06(-1)           && {\bf 1.12(-1)}     && {\bf 1.17(-1)} \\
\noalign{\smallskip}
CH$_3$OH  & 5.79(+4)     & 5.81(+4)  &&  {\bf 6.38(+4)}  && {\bf 5.56(+4)}     && {\bf 5.56(+4)}     && {\bf 5.56(+4)}     && {\bf 5.78(+4)} \\
\noalign{\smallskip}
CO$_2$    & 6.60(+3)     & 6.64(+3)  &&  9.50(-5)        && 9.06(+3)           && 9.06(+3)           && 9.06(+3)           && {\bf 6.84(+3)} \\
\noalign{\smallskip}
\hline
\end{tabular}
\end{center}
\end{table*}

\subsection{Full, two-particle competition}

Section 5 outlined a simple competition scheme that introduced an efficiency to the surface production rates. The scheme assumes that if one or other particle evaporates before reaction can occur, then the production process is over. However, this is an over-simplification; the final rate must take into account all opportunities for an individual, waiting particle to react before it is otherwise removed. Consider particle $A$ waiting on a grain for its reaction partner, $B$. Particle $B$ accretes onto the grain, and if $A$ and $B$ meet before evaporation of either species occurs, then they react. If particle $A$ itself evaporates before reaction can occur, then the process is over, as the part of the production rate associated with this scenario, $R_{form}(B) \cdot \langle N(A) \rangle$, presumes the presence of particle $A$ on the grain. But if particle $B$ evaporates, particle $A$ is still present and waiting for another particle $B$ to accrete, of which there is a steady stream. If particle $A$ can remain on the grain for another accretion timescale of $B$, without evaporating, then the reaction competition process may begin again. Two competition processes must therefore be considered: competition between reaction and evaporation of $A$ or $B$; and competition between accretion of $B$ and evaporation of $A$. Ignoring interference from other accreting particles, re-accretion may occur an arbitrary number of times. Thus, method A underestimates the reaction efficiency, particularly if reaction is slower than {\em one} of the evaporation rates.

Full, two-particle competition requires two efficiency factors, $\eta _{AB}(A)$ and $\eta _{AB}(B)$, depending on whether $A$ or $B$ is waiting on the grain. This gives a modified production rate:
\begin{eqnarray}
R_{mod}(AB) & = & R_{form}(B) \cdot \langle N(A) \rangle \cdot \eta _{AB}(A) \nonumber \\
            & + & R_{form}(A) \cdot \langle N(B) \rangle \cdot \eta _{AB}(B) .
\end{eqnarray}
To construct $\eta _{AB}(A)$, or equally, $\eta _{AB}(B)$, each individual competition process is considered. Firstly, the probability of reaction between $A$ and $B$ is defined in equation (17) as $C_{AB}$.

The probability that evaporation of $B$ takes place before reaction can occur, with $A$ remaining on the grain, is:
\begin{equation}
D_{AB}(A)=k_{evap}(B)/[k_{AB}+k_{evap}(A)+k_{evap}(B)] .
\end{equation}
In this case, $A$ has a chance to react with another arriving particle $B$. The probability that $A$ remains on the grain, avoiding evaporation, long enough for another particle $B$ to arrive is:
\begin{equation}
E_{AB}(A)=R_{form}(B)/[R_{form}(B)+k_{evap}(A)] .
\end{equation}
Hence, the probability that reaction does not occur at one opportunity, but that particle $A$ remains on the grain and has a chance to react with the next incoming particle $B$, is:
\begin{equation}
F_{AB}(A)=D_{AB}(A) \cdot E_{AB}(A)
\end{equation}
where $0 \geq F_{AB}(A) \geq 1$. For a waiting particle $A$, the rate associated with the first particle $B$ to arrive is, naturally, $R_{form}(B)$. However, if reaction only successfully occurs with the second particle $B$ to arrive, then particle $A$ should have waited a further accretion/formation period, $t_{form}(B)$. Hence, the arrival rate associated with reaction of $A$ with the second particle $B$ to arrive is $R_{form}(B)/2$. Similarly, for the third $B$ particle, the rate is $R_{form}(B)/3$, and so on.

Using the arrival rates and efficiencies defined above, the overall production rate in the case of particle $A$ waiting on the grain is constructed thus: \pagebreak
\begin{eqnarray*}
R_{mod,A}(AB) &=& R_{form}(B) \cdot \langle N(A) \rangle \cdot C_{AB} \\
               &+& \frac{R_{form}(B)}{2} \cdot \langle N(A) \rangle \cdot C_{AB} \cdot F_{AB}(A) \\
               &+& \frac{R_{form}(B)}{3} \cdot \langle N(A) \rangle \cdot C_{AB} \cdot F_{AB}^{2}(A) \\
               &\vdots& \\
               &=& R_{form}(B) \cdot \langle N(A) \rangle \cdot C_{AB}
                                       \sum_{n=0}^{\infty}\frac{F_{AB}^{n}(A)}{n+1}
\end{eqnarray*}
\begin{eqnarray*}
               &=& R_{form}(B) \cdot \langle N(A) \rangle \cdot C_{AB}\left\{ -\frac{1}{F_{AB}(A)}
                 \ln{\left[ 1-F_{AB}(A) \right] } \right\} .
\end{eqnarray*}
The efficiency is therefore defined as:
\begin{equation}
\eta _{AB}(A)=-\frac{C_{AB}}{F_{AB}(A)} \ln{ \left[ 1-F_{AB}(A) \right] }
\end{equation}
where $0 \geq \eta _{AB}(A) \geq 1$. As $F_{AB}(A) \rightarrow 0$, $\eta _{AB}(A) \rightarrow C_{AB}$. Calculation of $\eta _{AB}(A)$ and $\eta _{AB}(B)$ is trivial, and has a minimal effect on the run-time of the program.

Tables 4 -- 6 show, under the heading ``Method B'', the results of the application of these efficiencies to all reactions in the methanol system. In fact, the modification has a discernible effect only on abundances of CO and H$_2$CO, and then only in the low-density regime. Reactions involving these two species are hindered by activation energy barriers, making them uncompetitive compared to hydrogen evaporation; hence, their efficiencies are especially strongly affected by the change. Indeed, the abundances of CO and H$_2$CO for low density, using the modified rates, are now an exact match to the master-equation results, within computational accuracy. In the case of the intermediate-density regime, the abundances of CO and H$_2$CO are great enough that the modified rates exceed the standard deterministic rates, so equation (13) comes into effect. Hence, the CO and H$_2$CO rates are unaffected, whichever competition scheme is employed. In the high-density regime, $\langle N($CO$)\rangle, \langle N($H$_{2}$CO$)\rangle >>1$, putting them in the deterministic regime.

\begin{table*}
\caption[]{Modified rate results for high density, using input values from Stantcheva et al. (2002). Values in boldface show agreement within 10\% of the Monte Carlo or master-equation values.}
\label{tab6}
\begin{center}
\begin{tabular}{lrccccccccccc}
\hline
\hline
     \noalign{\smallskip}
Species   & \multicolumn{2}{c}{Stantcheva et al.} && Rate Eq. && \multicolumn{7}{c}{Modified Rates} \\
\noalign{\smallskip}
\cline{2-3}\cline{5-5}\cline{7-13}
\noalign{\smallskip}
          & Monte Carlo  & Master Eq. 23311 &&   && Method A    && Method B   && Method C          && Method D    \\
\noalign{\smallskip}
\cline{7-7}\cline{9-9} \cline{11-11}\cline{13-13}
\noalign{\smallskip}
          &              &          &&                   && ``Simple''         && ``Full''           && Method B +         && Method C +  \\
          &              &          &&                   && Competition        && Competition        && Poisson Prob.      && H-accretion \\
\noalign{\smallskip}    
\hline
\noalign{\smallskip}
H         & 0            & 8.29(-9)  &&  3.39(-10)       && 3.26(-9)           && 3.10(-9)           && {\bf 8.98(-9)}     && 7.37(-9) \\
\noalign{\smallskip}
O         & 1            & 5.76(-1)  &&  4.52(-1)        && 4.51(-1)           && 4.51(-1)           && 4.50(-1)           && 4.97(-1) \\
\noalign{\smallskip}
OH        & 1            & 5.97(-1)  &&  4.52(-1)        && 4.51(-1)           && 4.51(-1)           && 4.50(-1)           && 4.97(-1) \\
\noalign{\smallskip}
H$_2$     & 0            & 1.89(-6)  &&  8.63(-8)        && 7.52(-7)           && 7.15(-7)           && {\bf 2.07(-6)}     && 1.70(-6) \\
\noalign{\smallskip}
O$_2$     & 2.81(+5)     & 2.68(+5)  &&  {\bf 2.73(+5)}  && {\bf 2.72(+5)}     && {\bf 2.72(+5)}     && {\bf 2.71(+5)}     && {\bf 2.68(+5)} \\
\noalign{\smallskip}
H$_2$O    & 1.79(+5)     & 1.71(+5)  &&  2.49(+5)        && 2.25(+5)           && 2.26(+5)           && {\bf 1.81(+5)}     && {\bf 1.96(+5)} \\
\noalign{\smallskip}
CO        & 5.28(+5)     & 5.23(+5)  &&  6.43(+5)        && 5.93(+5)           && 5.94(+5)           && {\bf 5.14(+5)}     && {\bf 5.34(+5)} \\
\noalign{\smallskip}
HCO       & 0            & 1.53(-1)  &&  5.44(-3)        && 5.07(-2)           && 4.82(-2)           && 1.24(-1)           && 9.93(-2) \\
\noalign{\smallskip}
H$_2$CO   & 5.01(+4)     & 5.12(+4)  &&  1.50(+3)        && 2.21(+4)           && 2.16(+4)           && {\bf 4.63(+4)}     && 3.88(+4) \\
\noalign{\smallskip}
H$_3$CO   & 0            & 3.62(-2)  &&  2.61(-5)        && 4.10(-3)           && 3.79(-3)           && 2.37(-2)           && 1.63(-2) \\
\noalign{\smallskip}
CH$_3$OH  & 1.10(+4)     & 1.17(+4)  &&  4.81(+0)        && 1.89(+3)           && 1.82(+3)           && {\bf 1.16(+4)}     && 7.93(+3) \\
\noalign{\smallskip}
CO$_2$    & 5.82(+4)     & 6.01(+4)  &&  1.65(+3)        && 2.91(+4)           && 2.83(+4)           && 7.40(+4)           && {\bf 6.53(+4)} \\
\noalign{\smallskip}
\hline
\end{tabular}
\end{center}
\end{table*}

\subsection{Estimation of probabilities}

The modification method developed so far shows excellent agreement with the exact solutions, in the low-density regime; only O$_2$ and CO$_2$ show deviations greater than may be explained purely by computational rounding errors or minor differences in input values. Results for the intermediate-density regime are also very close, aside from O$_2$ and CO$_2$. Some populations that are less than unity are marginally different from the master-equation results; however, those cannot be regarded as wholly reliable, due to the use of cut-offs in the master-equation scheme. Other species with significant abundances are acceptably close to either the master-equation or Monte Carlo results.

In the high-density regime, the match is less acceptable; H$_2$, H$_2$O, H$_2$CO and CH$_3$OH all show significant deviations from the exact results. What causes these discrepancies?

In the intermediate- and high-density regimes, the average populations of some stochastic species get close to unity, even with the exact models. In the high-density case, O and OH average populations are around 0.5. This means that the probability approximation of equation (10) may no longer be accurate.

The probability that 1 or more atoms/molecules of species $i$ should be present on the grains at any arbitrary moment is:
\begin{equation}
P(i) \equiv \sum_{N=1}^{\infty}P_{N}(i)=1-P_{0}(i)
\end{equation} 
where $P_{N}(i)$ is the probability of population state $N(i)$. As $\langle N(i) \rangle$ approaches unity, the validity of equation (10) becomes questionable, as the probabilities of populations greater than 1 may become non-negligible. These probabilities are dependent on the degree of coupling between different population states.

To gauge the importance of the approximation, two extreme cases may be identified in the simple hydrogen-producing system in which the average abundance of atomic hydrogen would be less than 1. Firstly, consider a case in which $k_{H,H} >> R_{acc}($H$) >> k_{evap}($H$)$, i.e. H-accretion is much faster than H-evaporation, and the rate of the reaction H + H$\rightarrow$ H$_2$ is much faster than accretion. Here, evaporation is unimportant, as another H would accrete -- and reaction occur -- before evaporation should take place. The population state $N($H$)=2$ is short-lived, and so probabilites $P_{N>1}($H$)$ are negligible. In this case, the approximation $P(i) = \langle N(i) \rangle$ is valid in the entire range $0 \geq \langle N($H$) \rangle \geq 1$; although, in fact, the system reaches a steady-state value of $ \langle N($H$) \rangle =0.5$. The validity of equation (10) is the result of the strong coupling between the $N=2$ and $N=0$ states, engendered by the fast reaction between pairs of H-atoms.

Consider a second case, in which $k_{evap}(H) >> k_{H,H} >> R_{acc}(H)$. Here, evaporation dominates the destruction paths in all population states; hence, only population states that are adjacent are (strongly) coupled. Such a system may be understood as a queuing process of type M/M/$\infty$ (in Kendall notation); the population probabilities are therefore well described by a Poisson distribution. In this case, the probability of finding more than one particle of species $i$ at any instant is:
\begin{equation}
P(i)=1-\exp[- \langle N(i) \rangle ] .
\end{equation}
Thus, there are situations in this simple system where equation (10) is still valid, even when $\langle N(i) \rangle$ approaches unity; but there are others in which it is not -- equations (10) and (25) represent the extreme cases. It may be simplistically argued that, even for a large reaction network, $P(i)$ for any species with $\langle N(i) \rangle < 1$ should lie somewhere between these two extremes. The precise value of $P(i)$ would be dependent on the coupling between population states of the entire reaction network. 

Since the explicit evaluation of discreet population-state probabilities requires a master-equation approach, making an approximation to $P(i)$ is unavoidable. The simplest solution is to adopt one of equations (10) and (25). For a value $\langle N(i) \rangle =1$, these extreme cases differ in $P(i)$ by a factor of 1.58. In comparison to the errors inherent in chemical models, this difference is probably unimportant. However, one may argue that equation (25) is the more generally valid: In a complex network, reactions between many different species may occur, so it is generally less likely that destruction processes be dominated by reactions between like species, e.g. H + H $\rightarrow$ H$_2$. Reaction between heterogeneous particles would be more likely, so the potential for strong coupling between non-adjacent population states (in particular, $N(i)=0, 2$) should be mitigated.

The incorporation of equation (25) into the model is labelled ``Method C''. Tables 4 and 5 show that in the low- and intermediate-density cases, the change has only marginal, though beneficial, effects. However, the high-density results of Table 6 are much improved. The abundances of many important species, such as CO, CO$_2$ and H$_2$CO, show a very good level of agreement, whilst H$_2$O and CH$_3$OH are a perfect match to the exact results, within computational errors. The match for species with abundances less than 1 is also much improved, although not perfect. For high density, the results for all species are within 35\% of the exact results, and the majority are within 10\%.

\subsection{Hydrogen Accretion Competition}

Although method C produces an acceptable match to the exact methods in each density regime, certain species, most notably CO$_2$ and O$_2$, are not so well reproduced, particularly at low and intermediate densities. In fact, one further competition process must be considered to improve these results.

The formation of CO$_2$ and O$_2$ depend on the mobility of the oxygen atom on the grain surface. In comparison to atomic hydrogen, oxygen is very slow to diffuse between binding sites, due to its greater diffusion barrier, and its being much more massive than H, making tunnelling ineffective (although efficient tunnelling of hydrogen is also questionable, see Section 6). The consideration of competition up to now has concentrated solely on the case of two reactive species on a grain at any one time. However, the reaction rate for O-dependent reactions is so slow, at $\sim 4 \times 10^{-5}$ s$^{-1}$ (using values from Table 3), that a hydrogen atom may accrete before reaction occurs; it may then react with one or other of the reactants considered in the oxygen reaction. 

To test the influence of this competition process, terms equal to the accretion rate of hydrogen are inserted into equations (17), (20) and (21), for all reactions involving atomic oxygen. Reaction of atomic hydrogen with oxygen (or the other reactant, where applicable) is assumed to be instantaneous, if H-accretion takes place before the oxygen-dependent reaction can occur.

Tables 4 -- 6 show the results of this approach, labelled ``Method D''. In the low-density regime, the reproduction of the exact results is now perfect, within computational accuracy. In the intermediate density regime, the abundances of a number of species are now an exact match, or very close, e.g. H, H$_2$, H$_2$O, CO, HCO, H$_2$CO, CH$_3$O and CH$_3$OH, whilst O$_2$ is slightly less accurate than with method C. Even the worst match, CO$_2$, is much improved, falling within $10$\% of the exact result.

But in the high-density case, whilst some species such as CO show improved accuracy, many others diverge from the exact results, including formaldehyde (H$_2$CO) and methanol (CH$_3$OH). The level of agreement is generally worse than that achieved with method C. This may be explained by the fact that in this density regime, the accretion rate of oxygen itself is very fast. The accretion of hydrogen atoms interferes with certain reactions, but the further accretion of oxygen acts to mitigate this effect, as it may itself react with the newly accreted hydrogen. 

It becomes clear that in order to devise a generalised system to deal with accretion-competition effects, it would be necessary to consider not only the accretion of every reactive species, but also the probability that each might react with any other accreting species, in the context of a reaction between two entirely different reactants. Such a scheme could certainly be devised, but it would be extremely complex, both to implement and to fully understand. It would also, in all likelihood, be far more computationally expensive than, for example, Method C. Under those circumstances, the value of choosing such an approach over exact methods like the master equation would be questionable.

\subsection{Slow hydrogen regime}

In addition to the parameter set shown in Table 3, \cite{stant2} implemented the ``slow'' (M2) surface rates of Ruffle \& Herbst (2000). These are defined by diffusion barriers in line with the values suggested by \cite{katz} for the H + H $\rightarrow$ H$_2$ reaction. Implicit in these rates is the assumption that hydrogen-atom tunnelling through the diffusion barrier is inefficient; they represent a purely thermal hopping mechanism. Using these values, rate equations produce a perfectly acceptable match to the results of the Monte Carlo technique, in all regimes. Those same results are reproduced by even the simplest of the modified-rate methods presented here. The slower rates ensure either that the rate limit of equation (13) is reached, or that populations are high enough to place each reaction in the deterministic limit. This would not necessarily be the case in regimes with higher temperatures or smaller grains.

\begin{table}
\caption[]{Equations employed for each modification method}
\label{tabmethod}
\begin{center}
\begin{tabular}{lll}
\hline
\hline
\noalign{\smallskip}
Modification scheme  & Required Equations & Notes \\
\noalign{\smallskip}
\hline
\noalign{\smallskip}
Basic (H$_2$)       &  10 -- 13  & See Section 3.1 \\
\noalign{\smallskip}
Basic (continuous)  &  10 -- 15  &  \\
\noalign{\smallskip}
Method A            &  10 -- 18  &  \\
\noalign{\smallskip}
Method B            &  10, 13 -- 15, 17 -- 23  &  \\
\noalign{\smallskip}
Method C            &  13 -- 15, 17 -- 23, 25  &  \\
\noalign{\smallskip}
Method D            &  13 -- 15, 17 -- 23, 25  & See Section 6.3 \\
\noalign{\smallskip}
\hline
\end{tabular}
\end{center}
\end{table}

\section{Discussion}

It is clear that even a basic rate-modification scheme, such as applied here to the hydrogen system of \cite{barzel2}, is capable of achieving accurate results. However, application to the more complex water- and methanol-producing systems of Barzel \& Biham demonstrates the need to consider competition between reaction and evaporation processes. In fact, the influence of such competition is strongly dependent on the choice of surface parameters. The barriers to diffusion chosen by Barzel \& Biham are unusually high, around 85 -- 90 \% of the barrier against evaporation (see Table 2); this makes evaporation especially competitive with reaction. The lower barriers employed for the hydrogen-only system ($E_{b}$:$E_{D}\simeq0.69$, see Table 2) allow little scope for evaporation to compete; see Figure 1. It is for this reason that the basic modification scheme is so accurate for the H$_2$ system. Systems with yet lower $E_{b}:E_{D}$ ratios would also be very well reproduced with the basic method.

Indeed, the explicit treatment of evaporation--reaction competition in the methanol system of Stantcheva et al. (2002) makes only a small difference to the results, as evaporation is relatively slow. The greatest challenge in reproducing the low- and intermediate-density results presented by those authors is the treatment of reactions with activation-energy barriers. For such reactions, the competition comes not from evaporation but from the accretion of other species with which the reactants may preferentially react. This situation may be effectively treated by including the relevant accretion terms in the competition efficiencies; but, the approach is not universally applicable, as the high-density results demonstrate. However, method D is less necessary in that case, as method C already produces an acceptable match. In all regimes, method C still reproduces most abundances within 10\% of the exact results, with the remainder of species within a few tens of percent of their true values.

Whilst the neglect of accretion competition appears to make only a small difference to the results of the particular systems modelled here, would a change in parameters (e.g. accretion fluxes, diffusion barriers) lend a greater influence to such effects? Consider the formation of CO$_2$ and O$_2$, the only two species not accurately reproduced at low density by methods B and C. In that regime, formation of CO$_2$ occurs mainly by the reaction O + HCO $\rightarrow$ CO$_2$ + H, due to the high activation energy required for the alternative route. Also, the CO population is small, meaning that HCO is formed stochastically (see Section 5.1). Hence, the production rate of CO$_2$ should be $R_{acc}($O$) \cdot \langle N($HCO$) \rangle$; however, according to equation (13), the production rate may not exceed the deterministic maximum of $k_{O,HCO} \cdot \langle N($HCO$) \rangle \cdot \langle N($O$) \rangle$. The destruction of atomic oxygen is dominated by (stochastic) reaction with atomic hydrogen, so its population is given approximately by $\langle N($O$) \rangle = R_{acc}($O$) / R_{acc}($H$)$. As a result, the formation of CO$_2$ falls into the deterministic regime when $R_{acc}($H$) > k_{O,HCO}$. The implication for accretion competition is clear: hydrogen accretion cannot become more competitive than reaction, before the deterministic limit is reached and rate equations take over. The competition efficiency factor will thus never fall below $C_{O,HCO}=1/2$. A similar analysis yields $C_{O,O}=2/3$ for O$_2$ production. A parameter search that varies fluxes and diffusion barriers around their low-density values bears out these analyses: CO$_2$ and O$_2$ populations using method C vary from those of method D by no more than the stated factors, whilst populations of other species are an exact match. So far as the systems considered here are representative, method D may be considered unnecessary for the attainment of reasonably accurate results.

Some small inaccuracies are evident using method C in the high-density regime. A possible cause is a break-down in the two-particle assumption. For example, the average populations of O and OH in the high-density regime are both close to 0.5. Whilst equation (25) takes crude account of the probability of more than 1 of the same species being present on a grain at the same time, no account is taken of O and OH both being present. Such a possibility should lead to competition to react with, say, an accreting hydrogen atom. 

The Poisson probability of equation (25) is a simple approximation. It may be that the probability distibutions of, say, O and OH are skewed away from the $N=1$ state if an exact method is used. However, the full results of the hybrid master-equation technique employed by Stantcheva et al. (not shown here), indicate that the populations of stochastic species in the intermediate- and high-density regimes are very dependent on the cut-offs employed. The use of much larger cut-offs in their models could give populations in closer agreement with modified-rate method C. Method C does in fact allow for instantaneous populations, $N(i)$, of more than 1, through its assumption of a Poisson probability distribution -- equivalent to an infinitely large cut-off. Its main failure is in treating the more complex interactions between these heterogeneous populations. These failures are limited by the switch-over to deterministic rates when population states greater than 1 become probable.

Hence, in the intermediate- and high-density regimes, the only species that may be reliably compared between methods are those with large populations, which are treated accurately by the Monte Carlo method. With the exception of CO$_2$ (whose population discrepancies are explained above), all abundant species in the high-density regime are reproduced by method C to within 8\% of the Monte Carlo model values.

\section{Conclusions}

The production-rate modification schemes presented here substitute the deterministic production rates with so-called ``small-grain'' rates, applicable when populations are very small and reaction rates are fast. These basic rates are easily formulated, and have been more rigorously derived elsewhere using a master-equation approach (e.g. Lipshtat et al. 2004). A simple switching/rate-limiting system is employed for each reaction, allowing the deterministic rates to be used when appropriate. Even this simple scheme provides a highly accurate reproduction of the hydrogen system examined by Barzel \& Biham (2007a,b).

The consideration of competition between surface processes allows the water- and methanol-producing systems of Barzel \& Biham also to be very accurately reproduced. A further refinement is made to better estimate the probability of the presence of a reactant on the grains, improving accuracy in regimes in which average populations approach a value of 1. Excellent agreement with the results of the Monte Carlo models of Stantcheva et al. (2002) is achieved with this scheme, labelled method C.

The modification method is useful in that it trades off a small reduction in accuracy for a scheme that does not require the explicit treatment of population states. This allows the methods to be easily implemented in gas-grain chemical models. The emphasis on competition processes also renders the systems understandable in terms of physical processes, rather than probability distributions or other more abstract quantities.

The greatest practical obstacle to achieving a perfect reproduction of the results of exact methods is the existence of activation-energy barriers for certain key reactions. However, the method C results never fall more than a few tens of percent away from the true populations of species produced by these routes. Critically, the inaccuracies necessarily fall in a parameter space close to the stochastic--deterministic threshold, and are therefore limited by the switch-over to the standard rate-equations. Even these inaccuracies may be overcome by judicious consideration of competition processes; however, a comprehensive approach of this sort would be difficult, and computationally expensive. 

Perfect accuracy over all parameter space can of course only be achieved with exact techniques such as the Monte Carlo or master-equation methods. The fundamental cause of inaccuracies in the modification schemes is the break-down of the two-particle approximation that is applied, explicitly or implicitly, throughout this paper. The modification schemes, culminating in method C, act to diminish the effects of this break-down, thereby bridging more closely the stochastic and deterministic regimes.

It should be stated again that the purpose of the methods developed here is not to achieve perfect accuracy, but rather to achieve acceptable accuracy in the simulation of systems that cannot (currently) be treated using exact methods. In this aim, the modification schemes presented here appear promising. In many regimes, some of the modifications may not even be neccessary; competition between evaporation and reaction should only be important if large $E_{b}$:$E_{D}$ ratios are assumed, or if extremely small grains or extremely high temperatures are considered. The most important modification to the basic (continuous) method is that applied in method C. Even the use of just the basic (continuous) method would be an improvement over rate equations, in any regime. Table 7 details the equations necessary to implement each method.

Certain elements of the modification method are not rigorously tested by the systems used here, and should be further tested in future. This includes the switch-over to deterministic rates when populations become large; deterministic rates typically become valid before populations of 1 or more are achieved. Also, the ``full competition'' scheme first implemented in method B shows only a limited effect here. The accuracy of most simple reaction networks should not suffer greatly by its omission.

The methods of production-rate modification presented are simple enough that they may be implemented in significantly larger systems than those explored here, sufficient for a full treatment of gas-phase and grain-surface chemistry. In such networks it may also be necessary to treat photodissociation of surface species, being both a source of ``formation'' in equation (18), and a source of ``destruction'' for the evaluation of competition efficiencies in equations (17), (20) and (21). Photodissociation is, like evaporation, a deterministic process, rendering its implementation in the relevant equations trivial.

\begin{acknowledgements}
The author thanks the Alexander von Humboldt Foundation for a research fellowship. The author is grateful to B. Barzel and O. Biham for the communication of their model data, and to H. Cuppen and E. Herbst for critical readings of this paper.
\end{acknowledgements}


\begin{thebibliography}{}

\bibitem[Allen \& Robinson (1977)]{allen} Allen, M. \& Robinson, G. W. \apj, 212, 400

\bibitem[Awad et al. (2005)]{awad} Awad, Z., Chigai, T., Kimura, Y., Shalabeia, O. M. \& Yamamoto, T. 2005, \apj, 626, 262

\bibitem[Barzel \& Biham (2007a)]{barzel} Barzel, B. \& Biham, O. 2007a, \apj, 658, L37

\bibitem[Barzel \& Biham (2007b)]{barzel2} Barzel, B. \& Biham, O. 2007b, \jcp, 127, 144703

\bibitem[Biham et al. (2001)]{biham} Biham, O., Furman, I., Pirronello, V. \& Vidali, G. 2001, \apj, 553, 595

\bibitem[Caselli et al. (1998)]{caselli} Caselli, P., Hasegawa, T. I. \& Herbst, E. 1998, \apj, 495, 309

\bibitem[Caselli et al. (2002)]{caselli02} Caselli, P., Stantcheva, T., Shalabiea, O., Shematovich, V. I. \& Herbst, E. 2002, \planss, 1257, 1266

\bibitem[Cazaux et al. (2003)]{cazaux} Cazaux, S., Tielens, A. G. G. M., Ceccarelli, C., Castets, A., Wakelam, V., Caux, E., Parise, B. \& Teyssier, D. \apj, 593, L51

\bibitem[Chang et al. (2007)]{chang} Chang, Q., Cuppen, H. M. \& Herbst, E. 2007, \aap, 469, 973

\bibitem[Charnley et al. (1992)]{charnley92} Charnley, S. B., Tielens, A. G. G. M. \& Millar, T. J. 1992, \apj, 399, 71

\bibitem[Charnley et al. (1995)]{charnley95} Charnley, S. B., Kress, M. E., Tielens, A. G. G. M. \& Millar, T. J. 1995, \apj, 448, 232

\bibitem[Charnley et al. (1997)]{charnley97} Charnley, S. B., Tielens, A. G. G. M. \& Rodgers, S. D. 1997, \apj, 482, L203

\bibitem[Charnley (1998)]{charnley98} Charnley, S. B. 1998, \apj, 509, L121

\bibitem[Charnley (2001)]{charnley01} Charnley, S. B. 2001, \apj, 562, L99

\bibitem[Cuppen \& Herbst (2007)]{cuppen} Cuppen, H. M. \& Herbst, E. 2005, \mnras, 367, 1757

\bibitem[Duley \& Williams (1984)]{duley} Duley, W. W. \& Williams, D. A. 1984, Interstellar Chemistry (London: Academic)

\bibitem[Garrod \& Herbst (2006)]{garrod06} Garrod, R. T. \& Herbst, E. 2006, \aap, 457, 927

\bibitem[Garrod et al. (2008)]{garrod08} Garrod, R. T., Widicus Weaver, S. L. \& Herbst, E. 2008, \apj, 682, 283

\bibitem[Gould \& Salpeter (1963)]{gould} Gould, R. J. \& Salpeter E. E. 1963, \apj, 138, 393

\bibitem[Green et al. (2001)]{green} Green, N. J. B., Toniazzo, T., Pilling, M. J., Ruffle, D. P., Bell, N. \& Hartquist, T. W. 2001, \aap, 375, 1111

\bibitem[Hasegawa et al. (1992)]{hasegawa} Hasegawa, T. I., Herbst, E. \& Leung, C. M. 1992, \apjs, 83, 167

\bibitem[Hollenbach \& Salpeter (1971)]{hollenbach} Hollenbach, D. \& Salpeter E. E. 1971, \apj, 163, 155

\bibitem[Horn et al. (2004)]{horn} Horn, A., M{\o}llendal, H., Sekiguchi, O., Uggerud, E., Roberts, H, Herbst, E. Viggiano, A. A. \& Fridgen, T. D. 2004, \apj, 611, 605

\bibitem[Katz et al. (1999)]{katz} Katz, G. J., Furman, I., Biham, O., Pirronello, V. \& Vidali, G. 1999, \apj, 522, 305

\bibitem[Lipshtat et al. (2004)]{lipshtat} Lipshtat, A., Biham, O. \& Herbst, E. 2004, \mnras, 348, 1055

\bibitem[Rae et al. (2003)]{rae} Rae, J. G. L., Green, N. J. B., Hartquist, T. W., Pilling, M. J. \& Toniazzo, T. 2003, \aap, 405, 387

\bibitem[Ruffle \& Herbst(2000)]{ruffle} Ruffle, D. P. \& Herbst, E. 1998, \mnras, 319, 837

\bibitem[Shalabiea et al. (1998)]{shalabiea} Shalabiea, O. M., Caselli, \& Herbst, E. 1998, \apj, 502, 652

\bibitem[Stantcheva et al. (2001)]{stant1} Stantcheva, T., Caselli, P. \& Herbst, E. 2001, \aap, 375, 673

\bibitem[Stantcheva et al. (2002)]{stant2} Stantcheva, T., Shematovich, V. I. \& Herbst, E. 2002, \aap, 391, 1069

\bibitem[Tielens \& Hagen (1982)]{tielens} Tielens, A. G. G. M. \& Hagen, W. 1982, \aap, 114, 245


\end{thebibliography}
\end{document}